**Solidification and loss of hydrostaticity in liquid media used for pressure measurements.**

M. S. Torikachvili,[1] S. K. Kim,[2] E. Colombier,[2*] S. L. Bud'ko,[2] and P. C. Canfield[2]

[1] Department of Physics, San Diego State University, San Diego, California, 92182-1233, USA
00
[2] Ames Laboratory and Department of Physics and Astronomy, Iowa State University, Ames, Iowa, 50011, USA

**Abstract**

We carried out a study of the pressure dependence of the solidification temperature in nine pressure transmitting media that are liquid at ambient temperature, under pressures up to 2.3 GPa. These fluids are: 1:1 isopentane/n-pentane, 4:6 light mineral oil/n-pentane, 1:1 isoamyl alcohol/n-pentane, 4:1 methanol/ethanol, 1:1 FC72/FC84 (Fluorinert), Daphne 7373, isopentane, and Dow Corning PMX silicone oils 200 and 60,000 cst. We relied on the sensitivity of the electrical resistivity of Ba(Fe$_{1-x}$Ru$_x$)$_2$As$_2$ single crystals to the freezing of the pressure media, and cross-checked with corresponding anomalies observed in the resistance of the manganin coil that served as the ambient temperature resistive manometer. In addition to establishing the Temperature-Pressure line separating the liquid (hydrostatic) and frozen (non-hydrostatic) phases, these data permit rough estimates of the freezing pressure of these media at ambient temperature. This pressure establishes the extreme limit for the medium to be considered hydrostatic. For higher applied pressures the medium has to be treated as non-hydrostatic.

# 1. Introduction

Hydrostatic pressure (P) is a remarkable and versatile tool for experimental condensed matter physics. Pressure is a thermodynamic parameter, and therefore it has a bearing on phase stability. Phase diagrams having P as one of the variables are of great interest in many fields of research. By altering the interatomic spacings, P can affect the electronic, magnetic, structural, and optical properties in a number of important ways, which in turn have been very useful in the elucidation of novel phenomena. The proliferation of measurement techniques built around piston-cylinder, Bridgman, and diamond anvil pressure cells (DAC) has made several important discoveries possible. Hydrostatic pressure is most typically accomplished by immersing the sample in a pressure-transmitting medium (PTM), applying a force while constraining the medium, and locking the force in. The PTM can be a gas, liquid, or a finely pulverized solid at the time of loading, providing decreasing levels of hydrostaticity in that order. The choice of PTM depends largely on the requirements of the experiment and availability of instrumentation. It is important to emphasize that the level of hydrostaticity accomplished in the sample chamber can become even more of a concern as the temperature (T) and pressure change, and liquefaction, solidification, and changes in viscosity take place. Studies of hydrostatic limit under pressure in a number of PTM using various techniques have been carried out since the inception of high pressure research.[1-10] When cooling is required and the PTM of choice is a liquid at ambient temperature, mixtures are typically preferred over pure substances, in light of their lower freezing temperatures (Blagden's law) and therefore a wider T-range for hydrostaticity.

The non-hydrostatic condition can be very important in some experiments. For example, if a phase transition is driven sharply by pressure, non-hydrostaticity (i.e. a variance in pressure, or a convolution of hydrostatic and uniaxial components) introduces an experimental artifact such that the transition between the two phases is smeared out. A case in point is the observation of superconductivity (SC) in $CaFe_2As_2$,[11-12] $SrFe_2As_2$, and $BaFe_2As_2$ (Ref. 13) under the less-than-ideal hydrostatic pressure conditions of a frozen PTM. In the particular case of $CaFe_2As_2$, whereas SC with a $T_c \approx 12$ K could be observed under uniaxial pressure or pressure with an uniaxial component from a frozen PTM in a P-dome centered near 0.5 GPa,[11, 14] it was not observed under the high hydrostaticity condition of a He-gas pressure cell.[15] This discrepancy was caused by the combined effect of the extreme strain sensitivity in $CaFe_2As_2$ with the occurrence of a structural phase transition below the solidification temperature ($T_s$) of the PTM.[16] In particular, a structural transition from tetragonal to collapsed-tetragonal with a $\approx 9\%$ decrease and $\approx 2.5$ increase in the c- and a- lattice parameters, respectively, is stabilized by an hydrostatic P $\approx 0.4$

GPa at T ≈ 100 K. This large and anisotropic change in dimensions leads to poorly controlled stress and strain conditions when the sample is confined by a solid PTM, and it becomes structurally multi-phase. The portion of the high-temperature tetragonal phase which persists at low temperatures is responsible for the SC near ≈ 12 K.[16] Given such dramatic consequences of non-hydrostaticity of the PTM, it is of great relevance to know the pressure dependence of the solidification temperature of commonly used PTM. This $T_s(P)$ line delineates the separation between the unambiguous hydrostatic and the much more complex non-hydrostatic regions.

A recent thorough study of the pressure-induced solidification of 11 of the most commonly used gas (4) and liquid (7) PTM at ambient temperature was carried out by Klotz et al.[1] Their methodology consisted in inferring the freezing temperature of the PTM from the onset of a statistically measurable variance in the pressure at different locations of a DAC, as determined from the luminescence of ruby microspheres. Previous efforts to determine the solidification temperature of the 1:1 n-pentane/iso-pentane medium under pressure,[2-3, 17-18] yielded results similar to Ref. 1. More recently Kim et al. developed a methodology for identifying the solidification temperature of liquid PTM under pressure and determined $T_s$ vs $P$ for 1:1 isopentane/n-pentane and 4:6 light mineral oil/n-pentane for pressures up to 7.5 and 2.1 GPa, respectively.[7, 19] Their methodology relied on identifying small, anomalous, kink-like features in the ρ(T) data for $Ba(Fe_{1-x}Ru_x)_2As_2$ compounds (x = 0.21 in particular), that could be correlated with the solidification of the PTM. These features in ρ(T) always occurred at the same temperature for the same pressure, independently of the sample composition. Since these features moved in temperature as a function of pressure in a manner consistent with occurrence subtle changes in pressure conditions, and the feature at 300 K for the 1:1 isopentane/n-pentane PTM was consistent with the solidification data reported in the literature,[1] they took these features to represent the solidification of the PTM in the whole temperature range. Although these features are difficult to discern clearly in the ρ(T) data, they become readily seen in the derivatives dρ(T)/dT.[7, 19] The large sensitivity of the $AE$Fe$_2$As$_2$ (AE = alkaline earth) materials to the solidification of the PTM is consistent with the large anisotropy of the thermal expansion;[20] as the sample gets boxed in by the solid PTM it becomes subjected to stresses and strains due to the uneven changes in lattice parameters, and this is reflected in the behavior of ρ(T).

In this work, in addition to the data for 1:1 isopentane/n-pentane and 4:6 light mineral oil/n-pentane, which were included in Ref. 7, we determined the solidification temperature in pressures up to ≈ 2.3 GPa of five other liquid PTM: 1:1 n-pentane/isoamy alcohol, 4:1 methanol/ethanol, 1:1 Fluorinert FC72/FC84,

Daphne 7373, and pure isopentane. We also investigated Dow Corning PMX silicone oils of 200 and 600,000 centistoke (cst) viscosities, but were unable to unequivocally determine their freezing points.

## 2. Experimental Details

The Ba(Fe$_{0.79}$Ru$_{0.21}$)$_2$As$_2$ single crystals for this study were grown out of self-flux using a method described in Ref. 21. The measurements of $\rho(T, P)$ for pressures up to 2.3 GPa were carried out using a Be-Cu self-clamping piston-cylinder pressure cell, with a core of either hardened NiCrAl-alloy or non-magnetic tungsten carbide. The typical sample size was approximately 2.0 x 0.3 x 0.1 mm$^3$, the electrical resistance at ambient temperature was in the 100-500 mΩ range, and the excitation current (I$_{ex}$ ≈ 1 mA) was applied across the ab-plane. Four Pt leads were attached to the sample using Epotek H20E Ag-loaded epoxy. The sample leads, a coil of varnish insulated manganin wire, and a ≈ 10 mm length of bare Pb strip, which served as high-T and low-T manometers, respectively, were soldered to the tips of 12 Cu wires at the end of a Stycast-sealed feedthrough, all in a 4-wire configuration for resistance measurements. The manganin coil was wrapped around one of the copper leads of the feedthrough, such that its axis was approximately parallel to the loading force. This assembly was inserted in a PTFE cup filled with the PTM, and placed in the core of the cell. The anti-extrusion ring, pistons and lock nuts were positioned and the cell was closed. Force was applied and locked in at ambient temperature with a hydraulic press, using the manganin manometer as a reference. The actual pressure in this type of cell is known to change with temperature before stabilizing near ≈ 90 K (Ref. 22) due to the different dilation characteristics of the pressure cell constituents and PTM. The pressure at low temperatures was determined from the superconducting transition of the Pb manometer,[23] and was assumed to be the same for all T ≤ 90K. The pressure values between ambient temperature and 90 K were estimated by linear extrapolation between the 300 K and 90 K values yielded by the manganin and Pb manometers, respectively. The pressure cell was fit to a Quantum Design Physical Property Measurement System (PPMS-9), which monitored simultaneously the resistance of the sample, both manometers, and a calibrated Cernox sensor (CX-1030-SD) attached to the body of the cell, from which the sample temperature was inferred. The cell was first cooled with a sweep rate of ≈ 1-2 K/min, and then warmed up slowly at a rate of ≈ 0.35 K/min, which yielded a negligible temperature lag between sensor and sample. The $\rho(T, P)$ data for the T$_s$ vs P analysis were taken from the warm up cycle. In the course of this investigation we identified that anomalies in the $\rho(T)$ data for the manganin manometer correlated very well with the anomalies in the sample. Albeit smaller (the changes in d$\rho$(T)/dT were below 10% in magnitude compared to the sample), the manganin anomalies were valuable for cross-checking.

In some cases corresponding anomalies could be detected in Pb as well, though the resolution was poor. The T-range of the anomalies could be better resolved by inspecting the d$\rho$(T)/dT data. However, in light of the close T steps between the data points, in some cases there was a lot of scatter in the d$\rho$(T)/dT data, and we resorted to weighted smooth fits to carry out the analysis.

In addition to inferring the solidification of the PTM from the anomalies in the $\rho$(T) measurement on Ba(Fe$_{0.79}$Ru$_{0.21}$)$_2$As$_2$ and manganin, we set up a cooling station under a microscope in order to monitor the solidification of the PTM at ambient pressure. A PTFE capsule containing the PTM was inserted in a polyethylene foam disc which was placed inside of a beaker. Controlled cooling was achieved by adding small amounts of liquid nitrogen to the beaker. The temperature of the PTM was monitored with the tip of a chromel-alumel thermocouple which was immersed in the PTM. The vitrification temperature was determined visually, and a qualitative assessment of the viscosity was inferred from the resistance of the PTM to motion of the thermocouple.

It is noteworthy pointing out that some liquids do not fully solidify upon cooling, notably crude oils, and their freezing is better characterized by the pour point ($T_{pp}$), a temperature below which the flow characteristics are lost. Our analysis was not detailed enough to correlate the solidification of the PTM with a specific value of viscosity, although the $\rho$(T) features we track do seem to be associated with the onset of shears in a solidified medium.

## 3. Results and discussion

### 1- 1:1 isopentane/n-pentane

The $\rho/\rho_{300K}$ vs T data for Ba(Fe$_{0.79}$Ru$_{0.21}$)$_2$As$_2$ for three pressures in 1:1 isopentane/n-pentane PTM are shown in Fig. 1a, together with the derivatives d($\rho/\rho_{300K}$)/dT. The pressure values for T = 300 K and ≤ 90 K are indicated on the right side of the plot. The methodology for determining the solidification temperature of the PTM is indicated by the 2 arrows in the d($\rho/\rho_{300K}$)/dT curve for P = 0.77/0.25 GPa, and the value of $T_s$ (≈ 96.9 K) was taken as the midpoint. The ≈ 0.28 GPa pressure value corresponding to $T_s$ ≈ 96.9 K (Fig. 1b) was estimated by linear interpolation between the $P_{300K}$ ≈ 0.77 GPa (from manganin) and $P_{T≤90K}$ ≈ 0.25 GPa (from Pb) values. Although the shape of the anomalies in d($\rho/\rho_{300K}$)/dT varied for each PTM, we used the same methodology for determining $T_s$ and the corresponding solidification pressure $P_s$.

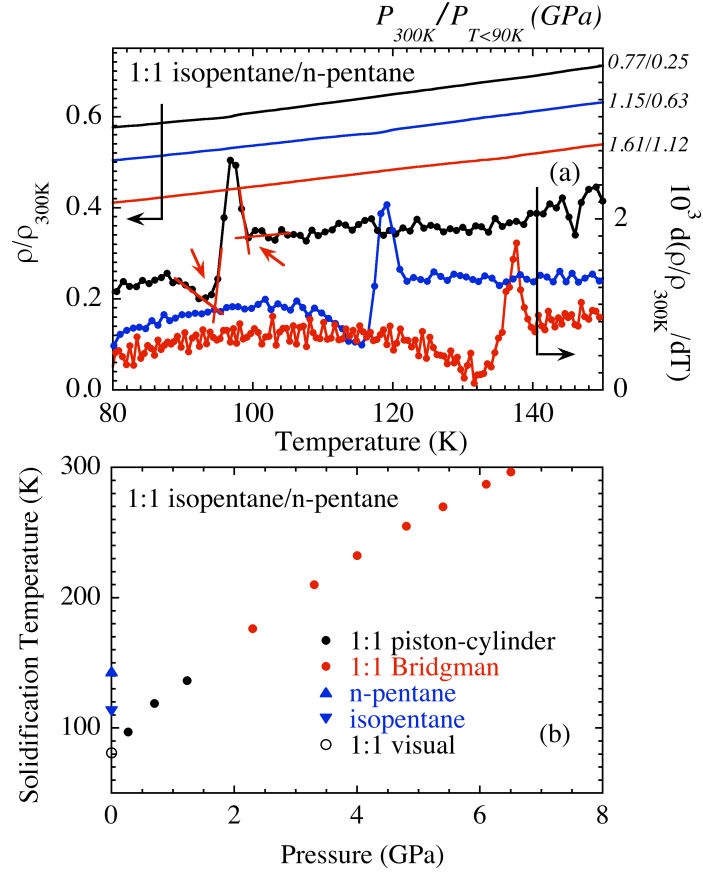

Fig. 1 – (color online) (a) Normalized electrical resistivity $\rho/\rho_{300K}$ (left) and $d(\rho/\rho_{300K})/dT$ (right) vs temperature for $Ba(Fe_{0.79}Ru_{0.21})_2As_2$ under pressure using a 1:1 isopentane/n-pentane mixture as the PTM. The curves are offset for clarity. The pressure values at 300 K and below 90 K, as determined from the manganin and Pb manometers, respectively, are shown to the right of the frame; (b) solidification temperature vs pressure determined from measurements in piston-cylinder (P < 2 GPa) and Bridgman (P > 2 GPa) cells. The methodology for determining the $T_s$ of the PTM is indicated by the arrows in the $d(\rho/\rho_{300K})/dT$ curve for P = 0.77/0.25 GPa, and taking the midpoint (see text). Pressure values at $T_s$ were estimated from a linear interpolation between the $P_{300K}$ and $P_{T≤90K}$ yielded by the manganin and Pb manometers, respectively. Approximate ambient pressure $T_s$ as determined by visual observation, as well as the values for pure n-pentane and isopentane are shown as well for reference.

A plot of $T_s$ vs P showing these data as well as the data in the 2 – 7 GPa range obtained with a Bridgman cell (Ref. 7) is shown in Fig. 1b. The value of $T_s$ for P = 0 was inferred from visual observation under the microscope. The solidification temperatures for pure n-pentane and isopentane at P = 0, taken from the material safety data sheets (MSDS) are also shown in Fig. 1b. The anomalies observed in the resistivity data for Ba(Fe$_{0.79}$Ru$_{0.21}$)$_2$As$_2$ correlate very well in temperature with anomalies seen in the resistance of the manganin (data not shown). The 1:1 isopentane/n-pentane PTM has low viscosity at ambient temperature. The visual observation upon cooling indicated some noticeable thickening near 130 K, and the viscosity kept increasing until the PTM vitrified near 81 K (the error bar is smaller than the size of the symbol in Fig. 1b).

**2- 4:6 light mineral oil/n-pentane**

The behavior of the normalized electrical resistivity $\rho/\rho_{300K}$ vs T for Ba(Fe$_{0.79}$Ru$_{0.21}$)$_2$As$_2$ under pressures up to 2.3 GPa in a 4:6 mixture of light mineral oil (Fisher Scientific CAS 8042-47-5) and n-pentane as the PTM is shown in Fig. 2a. The temperature derivatives are shown in Fig. 2b. In light of the scatter of the $d(\rho/\rho_{300K})/dT$ data, a weighted smoothing algorithm was used to determine the onset and completion of the anomalies more reliably. The values of $T_s$ shown in Fig. 2c were taken from the temperature midpoint between the beginning and end of the anomalies. The P = 0 value of $T_s$ for the 4:6 PTM (from visual observation), as well as the values for pure n-pentane (from the MSDS) and light mineral oil (from visual observation) are also shown in Fig. 2c. Similarly to the isopentane/n-pentane case, the anomalies in the resistivity due to solidification correlate very well in temperature with anomalies seen in the manganin manometer. The viscosity of the 4:6 oil/n-pentane is significantly higher than the 1:1 isopentane/n-pentane at ambient temperature. Upon cooling from ambient temperature we observed noticeable thickening below ≈ 210 K, and vitrification near 123 K.

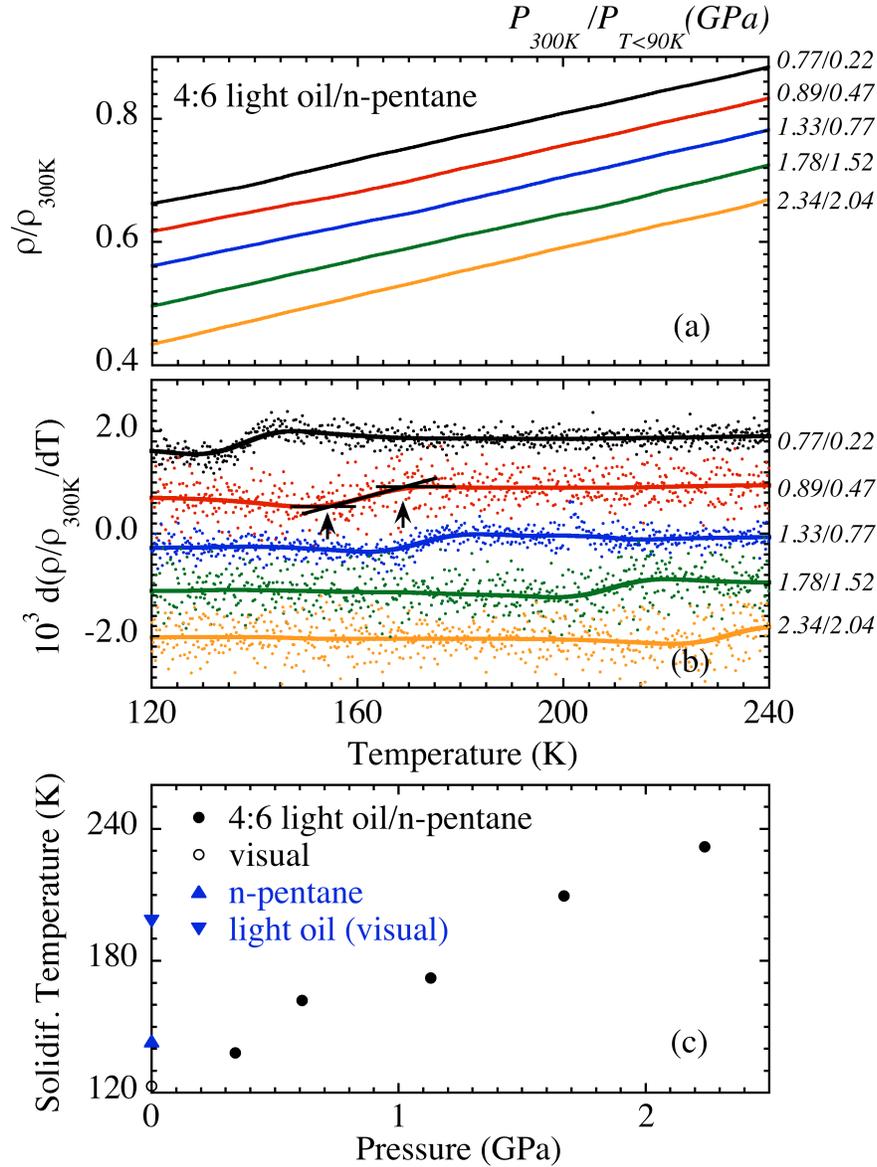

Fig. 2 – (color online) (a) Normalized electrical resistivity $\rho/\rho_{300K}$ vs temperature for Ba(Fe$_{0.79}$Ru$_{0.21}$)$_2$As$_2$ with a 4:6 light mineral oil/n-pentane mixture as the PTM; (b) derivatives $d(\rho/\rho_{300K})/dT$ for the data in the (a) pane. The solid lines are weighted smooth fits to the data. The curves in the (a) and (b) panes are offset for clarity; (c) solidification temperature vs pressure for P ≤ 2.3 GPa. Pressure values at $T_s$ were estimated from linear interpolation between the $P_{300K}$ and $P_{T≤90K}$ values (see text). The approximate ambient pressure $T_s$ (as determined from visual observation under the microscope), as well as the values for pure n-pentane and light mineral oil (visual) are shown as a reference.

## 3- 1:1 isoamyl alcohol/n-pentane

The behavior of the normalized electrical resistivity $\rho/\rho_{300K}$ vs T for Ba(Fe$_{0.79}$Ru$_{0.21}$)$_2$As$_2$ under pressures up to 2.0 GPa in a 1:1 mixture of isoamyl alcohol and n-pentane is shown in Fig. 3a. The resistivity anomalies correlated with the vitrification of the PTM are taken from the weighted fits of the d($\rho/\rho_{300K}$)/dT derivatives, as shown in Fig. 3b. These anomalies also correlate very well in temperature with anomalies seen in the resistance of the manganin manometer. The values of $T_s$ shown in Fig. 3c were taken from the temperature midpoint between the onset and completion of the anomalies. The P = 0 values of $T_s$ for the PTM (visual), as well as the values for pure n-pentane and isoamyl alcohol (from the MSDS) are also shown in Fig. 3c. The viscosity of 1:1 isoamyl alcohol/n-pentane starts to increase more noticeably below ≈ 210 K, and it vitrifies near 120 K.

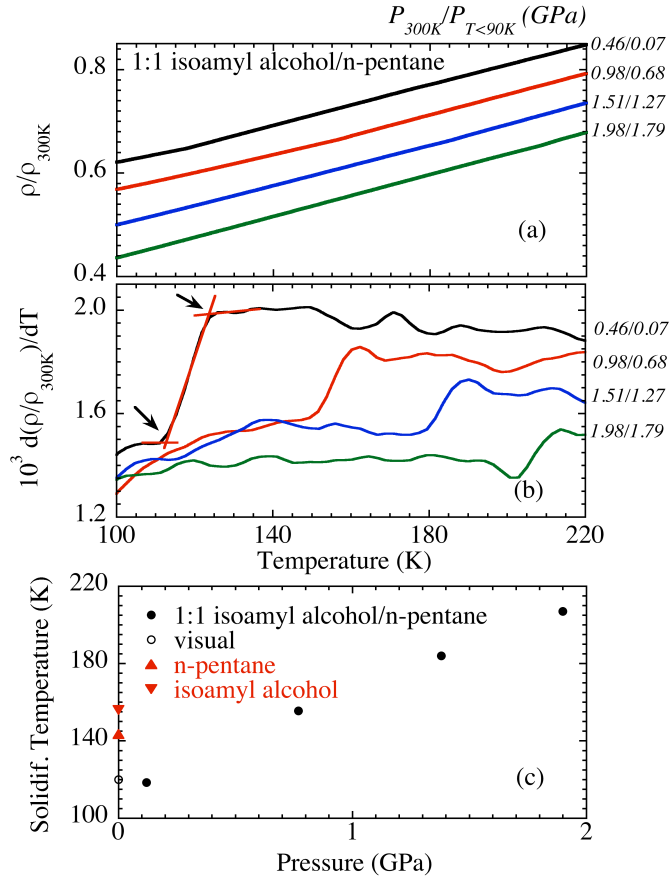

Fig. 3 – (color online) (a) Normalized $\rho/\rho_{300K}$ vs T; and (b) d($\rho/\rho_{300K}$)/dT vs T for Ba(Fe$_{0.79}$Ru$_{0.21}$)$_2$As$_2$ under pressure from 1:1 isoamyl alcohol/n-pentane PTM. The curves are offset for clarity. The density of point in (a) is high and there is no scatter in the data. The lines shown in (b) are weighted smooth fits to the data; (c) solidification temperature vs pressure.

## 4- 4:1 methanol/ethanol

The behavior of $\rho/\rho_{300K}$ vs T for Ba(Fe$_{0.79}$Ru$_{0.21}$)$_2$As$_2$ under pressures up to 2.0 GPa in a 4:1 mixture of methanol and ethanol is shown in Fig. 4a. The resistivity anomalies correlated with the vitrification of the PTM can be clearly seen in these data, as indicated by the arrows. The temperature derivatives of the resistivity are shown in Fig 4b, where the arrows indicate the methodology for determining the onset and completion of the anomalies. The corresponding anomaly in the resistivity of the manganin manometer at the highest pressure is shown as well. The values of T$_s$ vs P shown in Fig. 4c were taken from the midpoint between the onset and completion of the anomalies in $\rho$(T). The P = 0 values of T$_s$ for the PTM (visual), as well as the values for pure methanol and ethanol (from the MSDS) are also shown in Fig. 4c for reference. The T$_s$ (P=0) value came out surprisingly above an extrapolation of the T$_s$ vs P line to P = 0, suggesting that the vitrification observed visually is perhaps more revealing of a higher temperature pour point. The viscosity of 4:1 methanol/ethanol starts to increase very noticeably some 30 K above the vitrification point near 135 K.

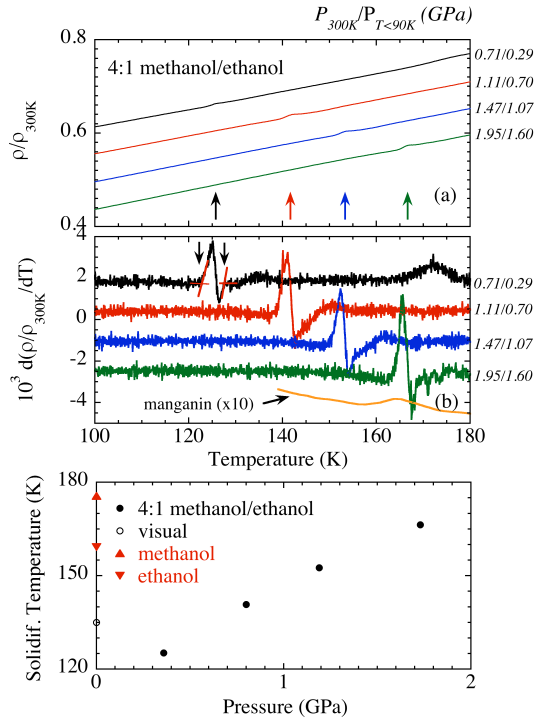

Fig. 4 – (color online) (a) Normalized $\rho/\rho_{300K}$ vs T; and (b) d($\rho/\rho_{300K}$)/dT vs T for Ba(Fe$_{0.79}$Ru$_{0.21}$)$_2$As$_2$ under pressure of 4:1 methanol/ethanol PTM. The curves in (a) and (b) are offset for clarity. The anomaly in the resistance of the manganin manometer for P$_{300K}$ = 1.95 GPa due to the vitrification is also shown in (b); (c) solidification temperature vs pressure.

## 5- 1:1: FC72/FC84

Fluorinert is a line of C-F-N fluorocarbon-based liquids manufactured by 3M. They are inert, stable, insulating, and have a number of properties that make them attractive to the electronic industry. Some of these properties make them attractive as PTM as well. In particular the absence of hydrogen makes them amenable to neutron scattering experiments under pressure. A thorough investigation of the hydrostatic limit at ambient temperature in pure and mixed FC's 70, 75, 77, 84, and 87 was carried out in Ref. 9, yielding limits in the ≈ 0.5 – 2.5 GPa range. In this work we focused on a 1:1 FC72/FC84 mixture. The behavior of the normalized electrical resistivity $\rho/\rho_{300K}$ vs T for Ba(Fe$_{0.79}$Ru$_{0.21}$)$_2$As$_2$ under pressures up to 1.6 GPa in a 1:1 mixture of FC72/FC84 is shown in Fig. 5a. The resistivity anomalies correlating with the vitrification of the PTM are subtle both in the Ba(Fe$_{0.79}$Ru$_{0.21}$)$_2$As$_2$ sample and the manganin manometer. They could still be identified in the temperature derivatives, as shown in Fig. 5b, though the vitrification features at higher pressures (data not shown) could not be determined unequivocally. The values of $T_s$ vs P shown in Fig. 5c were taken from the midpoint between the onset and completion of the anomalies. The P = 0 values of $T_s$ for the PTM (from visual observation), as well as the $T_{pp}$ values for pure FC72 and FC84 (from 3M's literature) are also shown in Fig. 5c for reference. The viscosity of the 1:1: FC72/FC84 PTM starts to increase more noticeably below ≈ 170 K, and vitrification takes place near 138 K.

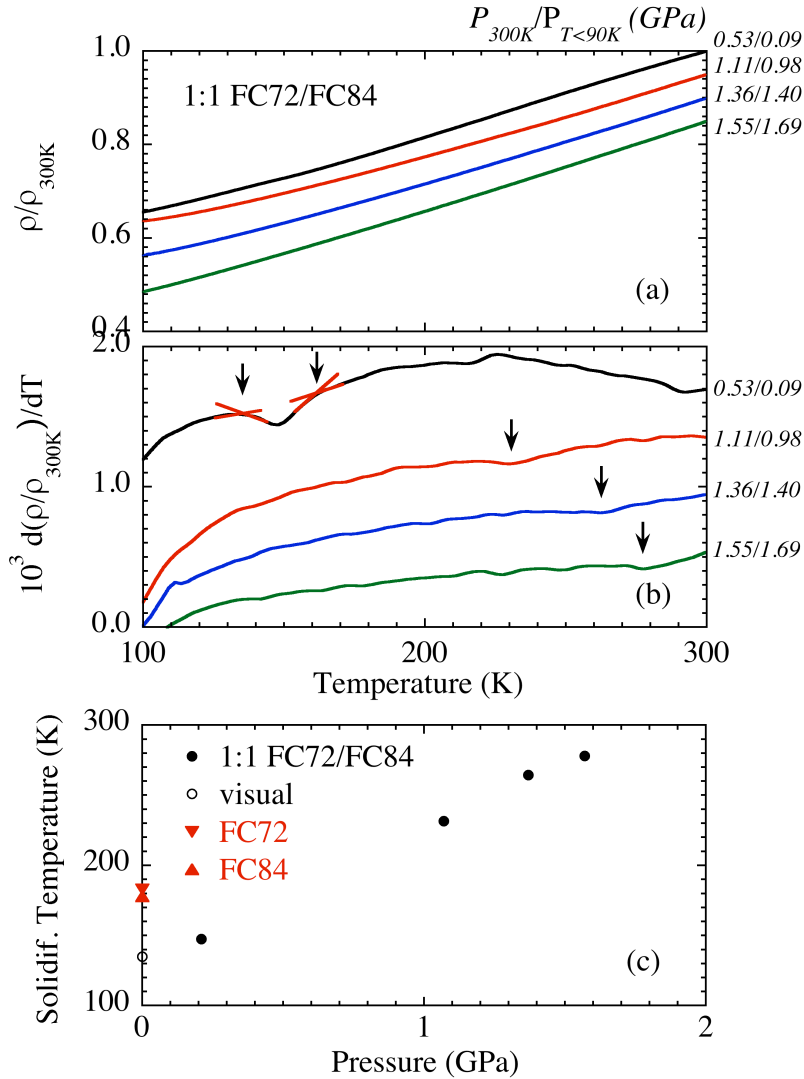

Fig. 5 – (color online) (a) Normalized $\rho/\rho_{300K}$ vs T; and (b) weighted smooth fit of $d(\rho/\rho_{300K})/dT$ vs T for Ba(Fe$_{0.79}$Ru$_{0.21}$)$_2$As$_2$ under pressure of 1:1 FC72/FC84 PTM. The curves are offset for clarity; (c) solidification temperature vs pressure. The P = 0 data shown for pure FC72 and FC84 are pour point temperature values taken from the 3M's product information. The $T_s$ value for the mixture was taken from visual observation under the microscope.

## 6- Daphne 7373

Daphne 7373 is a synthetic lubricant consisting of a mixure of olefin olygomers. It has been suggested that in piston-cylinder cells it has some advantage over other PTM because its change in pressure with temperature is less severe (≈ 0.15 GPa between 300 K and low temperatures for P ≤ 1 GPa), and nearly independent of pressure.[20] [Note: in the course of this work we found that the drop in pressure from 300 K to low temperatures in our cell was slightly higher and varied with pressure, e.g. ΔP ≈ 0.26 GPa for $P_{300K}$ ≈ 0.73 GPa, 0.20 GPa for $P_{300K}$ ≈ 1.14 GPa, and 0.17 GPa for $P_{300K}$ ≈ 1.51 GPa.]

The behavior of the normalized electrical resistivity $\rho/\rho_{300K}$ vs T for Ba(Fe$_{0.79}$Ru$_{0.21}$)$_2$As$_2$ under pressures up to 1.9 GPa in Daphne 7373 is shown in Fig. 6a. The resistivity anomalies which correlate with the vitrification of the PTM can be identified more clearly in the weighted smooth fits of the d($\rho/\rho_{300K}$)/dT vs T data, as shown in Fig. 6b. These signatures of vitrification can be identified as well in the derivatives d$\rho$/dT for the manganin manometer. The values of $T_s$ vs P shown in Fig. 6c were taken from the midpoint between the onset and suppression of the anomalies. The P = 0 values of $T_s$ estimated from the visual observation upon cooling is shown in Fig. 6c as well. The viscosity of Daphne 7373 at ambient temperature is significantly higher than the other organic PTM studied, and it starts to thicken noticeably ≈ 40 K above $T_s$.

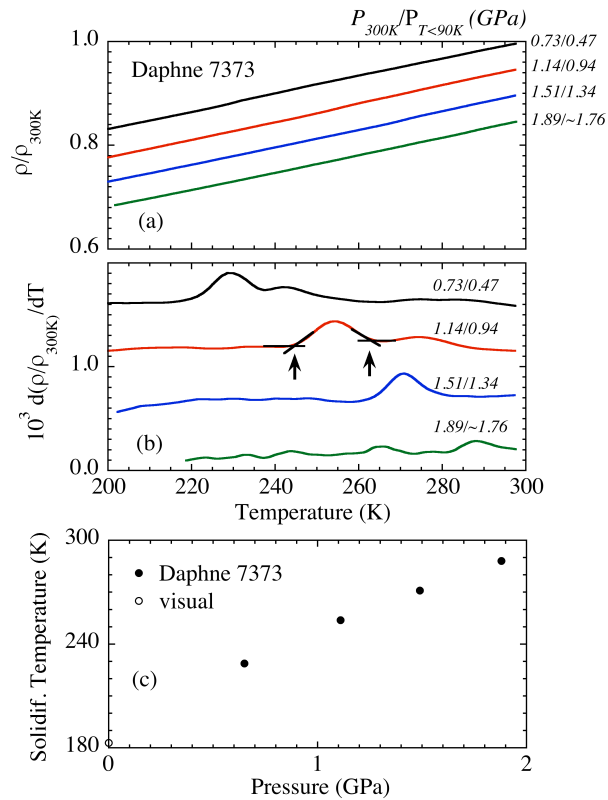

Fig. 6 – (color online) (a) Normalized $\rho/\rho_{300K}$ vs T; and (b) weighted smooth fits of d($\rho/\rho_{300K}$)/dT vs T for Ba(Fe$_{0.79}$Ru$_{0.21}$)$_2$As$_2$ with Daphne 7373 as the PTM. The curves are offset for clarity; (c) solidification temperature vs pressure. The P = 0 data point was taken from visual observation upon cooling.

## 7- isopentane

Displayed in Fig. 7 are the data corresponding to the use of pure isopentane as the PTM. In contrast to the mixed PTM and Daphne 7373, isopentane soldifies abruptly, without significant higher temperature thickening. The effect of freezing on the sample and contacts is severe as shown in the $\rho/\rho_{300K}$ vs T and $d(\rho/\rho_{300K})/dT$ vs T data of Figs. 7a and 7b, respectively. Actually the effect on the sample contacts is so severe that a more reliable determination of the onset and suppression of solidification features can be made by monitoring $d(\rho/\rho_{300K})/dT$ for manganin or Pb (Figs. 7c and 7d, respectively), since the latter are sturdier than the sample. A plot of $T_s$ vs P obtained from the midpoint of the anomalies in manganin is shown in Fig. 7e, together with the ambient pressure value of $T_s$. The values of $T_s$ under pressure determined from the Pb resistance are within 2 K of the values from manganin.

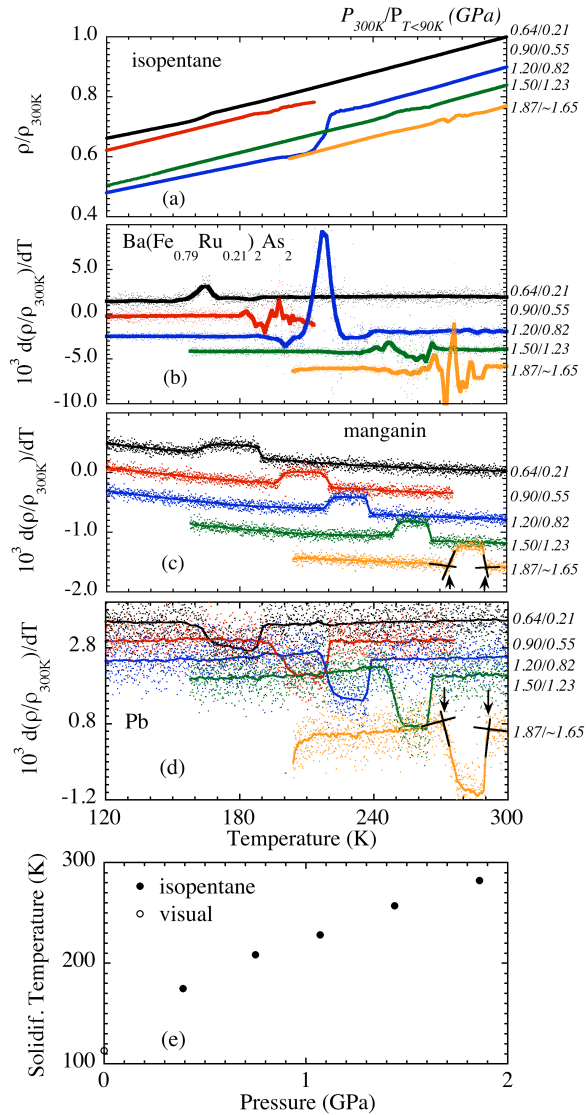

Fig. 7 – (color online) (a) $\rho/\rho_{300K}$ vs T; and (b) $d(\rho/\rho_{300K})/dT$ vs T for Ba(Fe$_{0.79}$Ru$_{0.21}$)$_2$As$_2$; (c) and (d) $d(\rho/\rho_{300K})/dT$ vs T (data and weighted fits) for the manganin and Pb manometers, respectively; and (e) $T_s$ vs P determined from the midpoint of the features in manganin. The curves in panels a-d are offset for clarity.

## 8-9- silicone oils

Silicone fluids have many desirable properties, and have been extensively used as PTM in high pressure studies. We tested Dow Corning PMX 200 and 60,000 cst as PTM for pressure measurements of the electrical resistivity of Ba(Fe$_{0.79}$Ru$_{0.21}$)$_2$As$_2$ in a piston-cylinder cell, and were unable to detect unambiguous features that corresponded to the vitrification of the medium. It is plausible that due to their higher viscosity the transition from hydrostatic to non-hydrostatic is too gradual and smears out the features in the sample and manganin. The pour points are 208 and 232 K respectively. Our measurements of R(T) in CaFe$_2$As$_2$ in pressures up to 0.7 GPa (data not shown) show significant smearing out of the tetragonal to orthorhombic transition (near 170 K at ambient pressure), and SC with a 12 K onset, both clear indications that neither fluid can be regarded as hydrostatic below 170 K.

## 4. Conclusions

In summary, we relied on the sensitivity of the electrical resistivity of Ba(Fe$_{0.79}$Ru$_{0.21}$)$_2$As$_2$ compounds to variances in pressure to map out the P-T phase diagrams separating the liquid (hydrostatic) and frozen (non-hydrostatic) phases of seven commonly employed PTM in high pressure experiments. These data were cross-checked with the correlated anomalies observed in the resistance of the manganin manometer which shared the pressure environment. The d$\rho$/dT data for manganin in Figs. 4 and 7 suggest a sensitivity to the solidification of the PTM 1-2 orders of magnitude lower than the sample. A summary plot of the T$_s$ vs P lines is shown in Fig. 8. Linear fit lines for P ≤2 GPa and extrapolations up to 300 K are shown as a reference for our discussion. An estimate of the solidification pressure at ambient temperature (P$_{s,\ 300K}$) can be made by extrapolation of the P-T curves to 300 K. However this has to be carried out with caution as extrapolations much beyond the actual range of the data can lead to large errors. Two cases in point are the 1:1 isopentane/n-pentane, and 4:1 methanol/ethanol PTM, in which our data can be compared to Klotz et al.[1] The 1:1 isopentane/n-pentane mixture was mapped out to ≈ 7 GPa (the data in the 0 – 2, and 2 – 7 GPa ranges were obtained with a piston-cylinder, and a modified Bridgman cell respectively). The T$_s$ vs P line for 1:1 isopentane/n-pentane is slightly sublinear as shown in Fig. 8, and it can be fit very well to a 2nd order polynomial, T$_s$ (K) = 87.12 + 42.66*P – 1.61*P$^2$ (P in GPa). This yields an ambient temperature solidification pressure value P$_{s,\ 300K}$ ≈ 6.7 GPa, in excellent agreement with the measurements of Klotz et al.[1] It important to point out that there is a ≈ 1.5 GPa discrepancy between the P$_{s,\ 300K}$ values measured and extrapolated linearly from the data below 2 GPa, underlining the shortcomings of an extrapolation much beyond the data range. In the case of the 4:1 methanol/ethanol PTM, a linear extrapolation of the T$_s$ vs P data from Fig. 8 to 300 K yields T$_{s,\ 300K}$ ≈ 6.2 GPa, which is ≈ 4.4 GPa below the P$_{s,\ 300K}$ ≈ 10.6 GPa value found in Ref. 1. Obviously the closer the data of the other PTM of Fig. 8 is to 300 K, the more reliable the linear extrapolation is likely to be.

The use of our methodology to map out the T$_s$ vs P phase diagrams for the two Dow Corning silicone oils (PMX 200 and 60,000 cst) lead to inconclusive results, and we were unable to map out their P-T phase diagrams reliably. However we were able to ascertain that below their $T_{pp}$ values they are distinctively non-hydrostatic.

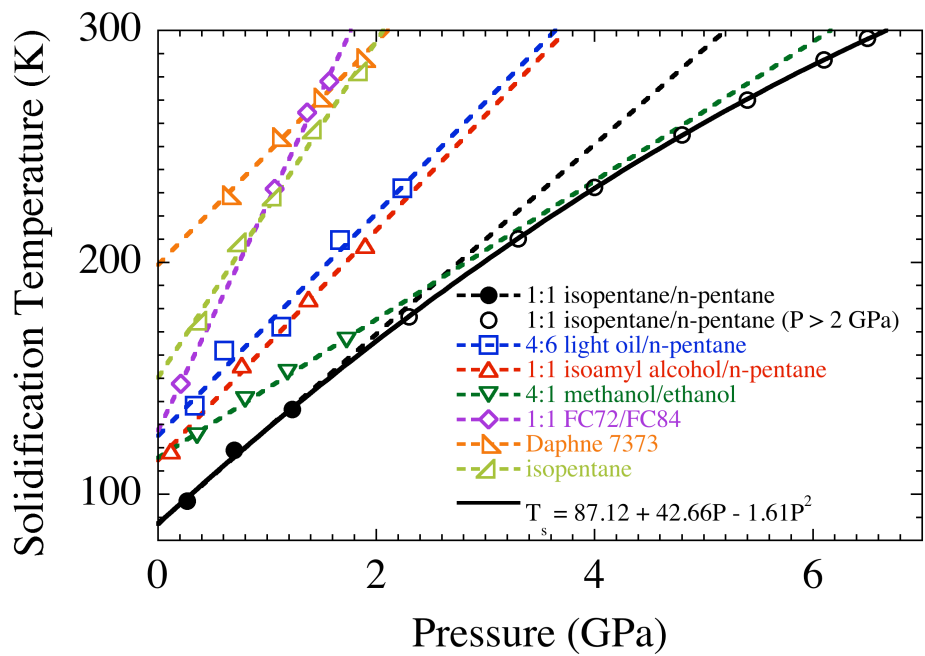

Fig. 8 – (color online) Solidification temperature vs pressure for 7 PTM. The lines are linear fits to the data for P < 2.3 GPa, extrapolated up to 300 K. The data for 1:1 isopentane/n-pentane is fit to a 2nd order polynomial; the discrepancy between the $P_{s,\,300K}$ values between the linear extrapolation from low P and the actual value is ≈ 1.5 GPa.

**Aknowledgments**

The authors are grateful to Dr. Alexander Thaler for participating in the synthesis of the samples for this research. Support from NSF Grant No. DMR-0805335 for the work at SDSU is gratefully acknowledged. Work at the Ames Laboratory was supported by the US Department of Energy, Basic Energy Sciences, Division of Materials Sciences and Engineering under Contract No. DE-AC02-07CH11358